\documentclass{elsart}

 \usepackage{graphics}
 \usepackage{graphicx}
 \usepackage{epsfig}

\usepackage{amssymb}

\newcommand{\be}{\begin{equation}}
\newcommand{\ee}{\end{equation}}
\newcommand{\bfig}{\begin{figure}}
\newcommand{\efig}{\end{figure}}
\newcommand{\bea}{\begin{eqnarray}}
\newcommand{\eea}{\end{eqnarray}}

\newcommand{\raa}{$R_{AA}$ }

\newcommand{\eqnraa}{R_{AA}}

\newcommand{\rhopartcomma}{$\rho_{\mathrm{part}}$}
\newcommand{\eqnrhopart}{\rho_{\mathrm{part}}}

\newcommand{\eqnnpart}{N_{\mathrm{part}}}

\newcommand{\dnslashdy}{$dN_g/dy$ }
\newcommand{\dnslashdycomma}{$dN_g/dy$}
\newcommand{\eqndnslashdy}{dN_g/dy}
\newcommand{\as}{\alpha_s}
\newcommand{\alphas}{$\as$ }

\newcommand{\qhatcomma}{$\hat{q}$}

\newcommand{\eqnlinput}{L_{\mathrm{input}}}

\newcommand{\eqnrhojet}{\rho_{\mathrm{Jet}}}

\newcommand{\pt}{$p_\perp$ }

\newcommand{\eqnpt}{p_\perp}

\newcommand{\raapt}{$\eqnraa(\eqnpt)$ }

\newcommand{\fig}[1]{Fig.~\ref{#1}}

\begin{document}

\begin{frontmatter}



\title{Overcoming Fragility}


\author{William Horowitz}
\ead{horowitz@phys.columbia.edu}

\address{Columbia University, 538 West 120$^{th}$ Street, New York, NY 10027, USA}

\begin{abstract}
We examine the sensitivity and surface bias of convolved DGLV radiative and elastic loss in a realistic geometry including Bjorken expansion.  We find that this more faithful treatment of the medium density is not {\em a priori} reproducible via fixed length approximations, and neither the fragility nor the surface emission of BDMPS-based models is seen.
\end{abstract}

\end{frontmatter}

\section{Introduction}
The ultimate goal of hard probe physics is jet tomography, the use of the high-\pt hadronic attenuation pattern to learn about the medium density.  In order to believe that this is an achievable goal one needs a theoretical description of the processes involved that describes the data and that has its parameter(s) relatable to the matter density under calibrated control.  

For these tools to be truly tomographic, however, there is the additional requirement that the resulting medium measurement be reasonably precise.  The overall precision of density determination is set by a combination of experimental precision and theoretical sensitivity: if the experimental data are infinitely precise, any model with the previous paragraph's characteristics could asses the medium density with similar precision; but experimental error bands exist and must be taken seriously, limiting the precision of medium determination by the innate sensitivity of the theoretical model to changes in the aforementioned parameter(s).  As time progresses experimental error bars tend to decrease, allowing for a more precise measurement.

\cite{oldmodels} claimed that BDMPS-based radiative energy loss models are a fragile probe of the medium; i.$\,$e.~large changes in \qhatcomma, the input parameter for the theory, are not reflected in \raa predictions incompatible with data.  Surface emission, the process in which only the partons produced very close to the medium edge escape for observation, was posited as the reason for their observed insensitivity to increased medium opacity.  This is an appealing argument, and the latter two of \cite{oldmodels} provided evidence for this idea.  The very pessimistic conclusion, then, was that jet tomography is not possible at RHIC and will not be possible at the LHC.  

We counter the dire claims in \cite{oldmodels} by first noting that several approximations were made in their description of the medium produced in heavy ion collisions.  In the first of \cite{oldmodels}, the authors used a hard cylinder initial nuclear density and did not include Bjorken expansion.  The latter works of \cite{oldmodels} used the realistic Woods-Saxon nuclear density distribution but again neglected Bjorken expansion.  These certainly aided in alleviating some of the high numerical cost involved in the calculations.  However we feel these geometrical simplifications strongly biased their results to the surface of the medium resulting in a significant loss in model sensitivity.  Moreover, experimental error bars have shrunk considerably in the time since \cite{oldmodels} were published. These issues raise serious doubts about the validity of their conclusions.  In this paper we will explore the sensitivity and surface bias of a theoretical model based on radiative and elastic energy loss in a realistic diffuse medium that includes Bjorken expansion.

\section{Results}
\begin{figure}[h]
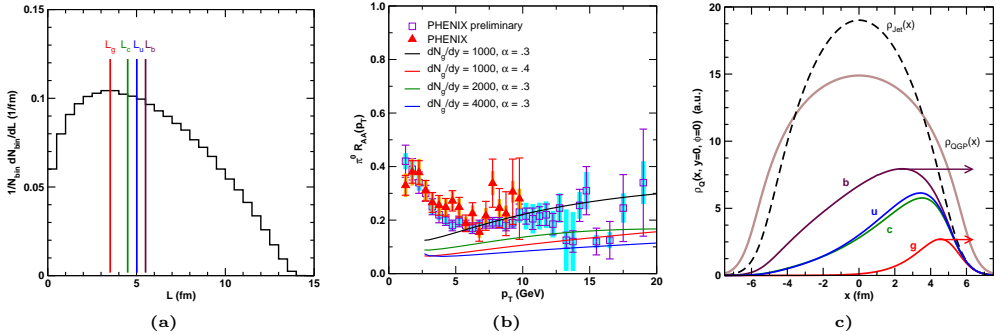

\vspace{.15in}
\begin{center}
$\begin{array}{c@{\hspace{.15in}}c@{\hspace{.15in}}c}
   \epsfig{file=pofl4.eps,width=.3\columnwidth,angle=0} & 
   \epsfig{file=060605pionfragility.eps,width=.3\columnwidth,angle=0} &
   \epsfig{file=Survivalplot_v5.eps,width=.29\columnwidth,angle=0} \\ [-.17in]
{\mbox {\tiny {\bf (a)}}} & {\mbox {\tiny {\bf (b)}}} & {\mbox {\tiny {\bf c)}}}
\end{array}$
\end{center}
\vspace{-.1in}
\caption{{\footnotesize All plots depict midrapidity for most-central collisions. (a) Distribution of path lengths traversed by hard scatterers; a purely geometric quantity, it is the same for all jet varieties. Note that no {\em single} pathlength can reproduce the results of the full distribution for all parton flavors. (b) Model sensitivity to varying \dnslashdy, which is much greater than observed in \cite{oldmodels}.  Susceptibility to changes in \alphas suggests the need for a relaxing of the fixed coupling constant approximation. (c) Transverse coordinate
$(x,0)$ distribution of surviving
$\eqnpt=15$ GeV, $Q=g,u,c,b$ jets moving in direction $\phi=0$
as indicated by the arrows. Units are arbitrary for illustration.
The binary collision distribution
of initial jet production points, $\rho_{\rm Jet}(x,0)$, is shown; the ratio $\rho_Q/\rho_{\rm Jet}$ gives the local quenching factor
including elastic and inelastic energy loss though
the bulk QGP matter distributed as $\rho_{\rm QGP}(x,0)$.}}
\label{fig1}
\end{figure}

Unfortunately, the use of realistic geometry is nontrivial.  The analytical radiative energy loss calculations necessarily were made in simpler geometries, and the elastic energy loss involves a line integral through the local temperature field of the medium; a mapping from the diffuse, \rhopartcomma-like medium into the approximate, idealized geometrical settings in which the theoretical calculations were made is required.  For the WHDG model using convolved elastic and inelastic energy loss, we define the parton's path ``length" as $\eqnlinput(\vec{x},\hat{n};b) = \int dl \eqnrhopart(\vec{x}+\hat{n}l;b)/\langle \eqnrhopart \rangle (b)$, where $\langle \eqnrhopart \rangle (b) = \int dxdy \eqnrhopart^2(x,y;b)/\eqnnpart(b)$.  The probability of a parton having a particular path length is then given by $P(L;b) = \int dxdyd\phi T_{AA}(\vec{x};b) \delta(L-\eqnlinput(\vec{x},\hat{n};b)/2\pi\eqnnpart (b)$.  We consider a diffuse Woods-Saxon participant density created by the Glauber profiles and include 1D Bjorken expansion.  \fig{fig1} (a) shows the wide distribution $P(L)$ of input path lengths resulting from a most central collision.  We also find the flavor-dependent single lengths $L_Q$, $Q=g,u,c,b$ that best reproduce the energy loss resulting from the full $P(L)$ distribution; these are represented by the vertical lines in \fig{fig1} (a). What is most striking in \fig{fig1} (a) is the hierarchy of $Q$-dependent length scales.  No single, representative path length can account for the full distribution of lengths, and their hierarchy of lengths is not {\em a priori} obvious.  

A detailed explanation of our DGLV radiative convolved with elastic energy formulation can be found in \cite{Wicks:2005gt}.  The single input parameter is \dnslashdycomma, which sets the normalization scale between the actual medium density and \rhopartcomma.  We mention briefly that, as discussed in \cite{Wicks:2005gt}, we have a fixed \alphas coupling.  One can see from \fig{fig1} that our model agrees with the data for a medium density compatible with the observed entropy constraint, $\eqndnslashdy\sim1000$.  Using fluctuating path lengths in a realistic and expanding background medium geometry, the predicted \raa is no longer consistent with data when the medium density is artificially enhanced by a factor of two.  Our sensitivity to medium density changes is enhanced over pure radiative loss models by the inclusion of elastic energy loss, due to its smaller width for fluctuations relative to radiative  \cite{Wicks:2005gt}.  We note, however, that models using GLV-type radiative loss or higher-twist loss only are similarly not fragile \cite{Vitev:2005ch}.  We therefore conclude that the pion \raapt is in fact quite a good experimental signal for jet tomographic studies at RHIC.

We also do not observe surface emission in our model of RHIC physics.  Define a parton's survival probability, the chance that a parton produced at a specific location and making an angle $\phi$ with respect to the reaction plane is observed escaping the medium, as $\rho_Q(\vec{x},\phi)=\eqnrhojet (\vec{x})\int d\epsilon (1-\epsilon)^n P_Q(\epsilon;L(\vec{x},\phi))$,
see \fig{fig1}. Again one sees that no single, representative path length can account for the distribution of all flavors. While all are biased to the surface, a necessary consequence of nonzero energy loss, none of the distributions can be categorized as surface emission. The characteristic widths of these distributions range from $\Delta x \approx 3-6$ fm.  Similar results of partonic survival from deep within the medium are seen in more careful BDMPS-based studies that use realistic geometry and Bjorken expansion and the higher-twist energy loss model, \cite{Renk:2006pw}.  We therefore conclude that the observed pions do explicitly probe significant portions of the medium.

\section{Conclusions}
In order to make strong statements on fragility, energy loss models must use realistic nuclear densities with Bjorken expansion.  Otherwise the naive explanation of model insensitivity to increased medium density, surface emission, is a reflection of the surface bias of the oversimplified base geometry, not of the actual surface tendencies of observed high-\pt partons.  Unfortunately, the use of realistic geometries is nontrivial; RHIC is not a brick.  Our model for high-\pt attenuation with convolved elastic and inelastic energy loss in a realistic geometry with Bjorken expansion is quite sensitive to changes in medium density and does not display the properties of surface emission.  

Further work is needed: as evidenced by \fig{fig1}, \raa depends strongly on the coupling; the assumption of fixed \alphas must be relaxed.  The first reference of \cite{Renk:2006pw} showed that several trivial unphysical models approximated the nearly flat in \raa data; we must show that our model of radiative, elastic, and path length fluctuations can reproduce the plethora of high-\pt observables.  If we convince ourselves that our theoretical description is correct, then we may return to the precise experimental \raa measurement to tomographically probe the medium.




\end{document}